\documentclass[article]{article}
\usepackage{authblk}
\usepackage{graphicx}
\usepackage{amsmath}
\usepackage{mathtools,slashed}
\usepackage[utf8]{inputenc} 
\usepackage[T1]{fontenc}
\newcommand{\RNumb}[1]{\uppercase\expandafter{\romannumeral #1\relax}}
\usepackage{braket}
\usepackage[11pt]{extsizes}
\usepackage{tikz}
\usepackage[left=2.8cm,right=2.8cm,
    top=3.0cm,bottom=3.3cm,bindingoffset=0cm]{geometry}
    
\usepackage[T1]{fontenc} 
\usepackage{amsfonts}
\usepackage{xcolor}
\usepackage{hyperref}
\definecolor{linkcolor}{HTML}{FF0000} 
\definecolor{urlcolor}{HTML}{799B03}
\begin{document}
\title{\textbf{Grassmannian and Flag sigma models on interval:
phase structure and L-dependence}}
\author{D. Pavshinkin\thanks{email: dmitriy.pavshinkin@phystech.edu}\\
\small{{\textit{Moscow Institute of Physics and Technology, Dolgoprudny 141700, Russia}}\\
\textit{Institute of Theoretical and Experimental Physics, Moscow 117259, Russia}}}
\date{}

\maketitle              
\begin{abstract}
We discuss the two-dimensional Grassmannian $SU(N)/S(U(N-2)\times U(2))$ and the flag $SU(N)/S(U(N-2)\times U(1)\times U(1))$ sigma models on a finite interval and construct analytical solutions of gap equations in the large N limit. We show that the flag model admits a homogeneous solution for ``mixed'' Dirichlet-Neumann (DN) boundary conditions only for sufficiently large length $L$  and undergoes a phase transition from the phase of partly broken gauge symmetry ($U(1)$) to the symmetric phase ($U(1)\times U(1)$) for large $L$. On the other hand, the Grassmannian model has a detached phase with one massive and one massless non-zero condensates that completely break $U(2)$ gauge symmetry. This phase lives on a region of $L$ bounded from above and has to use the Robin boundary conditions. We also examine the L-dependence of the total energy and detect the linear growth inherent to confining string in all phases. 

\end{abstract}
\section{Introduction}
\label{sec:1}
As it can be seen from a lot of works on QCD-like theories a phase structure, geometry and  vacuum have nontrivial influence on each other. A review of non-perturbative Casimir effects in theories possessing mass gap, confinement and chiral symmetry breaking phenomena was recently done in \cite{Chernodub}. It was natural to probe these problems for the well-studied $\mathbb{CP}^{N-1}$ non-linear sigma model (NLSM). This toy model has asymptotic freedom, dynamical mass generation \cite{D'adda,Witten,Novikiv} and may undergo the Higgs-Coulomb phase transition on the interval in the large-N limit for special boundary conditions \cite{Milekhin.1,GSY2006,Monin,Milekhin.2}. 

In this paper, we continue the study of homogeneous solutions for non-supersymmetric 2d NLSMs with more general target spaces, Grassmannian $Gr(N,2)= SU(N)/S(U(N-2)\times U(2))$ (\cite{Pisarski}) and flag $F(N,2,1)= SU(N)/S(U(N-2)\times U(1)\times U(1))$ \cite{Helgason}), that was started in \cite{Pavshinkin}. Since these manifolds have positive Ricci curvature, the corresponding field theories are asymptotically free. Also there is dynamical mass generation \cite{Itoh.Oh.Ryou1,Itoh.Oh.Ryou2}. The phase structure of the compactified 2d $SU(3)/S(U(1)\times U(1))$ flag sigma model with respect to $\theta$-terms was considered in \cite{Hongo}. Study of instanton-like solutions in the Grassmannian sigma model on $\mathbb{R}\times S^1$ was carried out in \cite{Misumi,Eto}.
Like the $\mathbb{CP}^{N-1}$ NLSM, these models are effective theories describing orientational moduli on the worldsheet of non-Abelian strings \cite{Tong,GSY2005}. The flag sigma models also appear in the low-energy description of anti-ferromagnetic spin chains \cite{Bykov1,Wamer,Bykov2}. The 1/N-expansion for the model with $SU(N)/S(U(N-2)\times U(1)\times U(1))$ target space was constructed in \cite{BykovN}. In the field of view of this work is the study of the phase structure of this sigma model in the large-$N$ limit on the finite interval.

 In section~\ref{sec:2} we construct the Lagrangian for the simplest flag sigma model $F(N,2,1)$. In general, there are two different coupling constants. Therefore two different dimensional parameters, $\Lambda_1$ and $\Lambda_2$ (e.g. $\Lambda_1\geq\Lambda_2$), are generated. The theory passes into $Gr(N,2)$ if the coupling constants coincide. Then we get the effective action by integrating out all but two fields. The gap equations for the theory on the interval are constructed in section~\ref{sec:3}.
Section~\ref{sec:4} is the main part of this paper. Here we investigate the phase structure of the models. For $F(N,2,1)$ model the region of $L>\Lambda_2^{-1}$ corresponds to unbroken $U(1)_A\times U(1)_B$ gauge symmetry. For $\Lambda_2^{-1}>L>\Lambda_1^{-1}$ we are in phase with $U(1)_A$ symmetry. The quantity $\Lambda_1^{-1}$ plays the role of ultraviolet cutoff (or string thickness) since for $L<\Lambda_1^{-1}$ there is no solution of the gap equations, and the gauge symmetry can not be broken completely.
On the other hand, when the coupling constants coincide and $F(N,2,1)$ turns to $Gr(N,2)$, for special choice of boundaries we have a solution only for $\Lambda_1^{-1}>L>0$. This region corresponds to the phase with completely broken $U(2)$ gauge symmetry.
In Section~\ref{sec:5}, we examine the L-dependence and find out that the linear term corresponding to the string tension appears in all phases. Also we analyze the L\"{u}scher term that appears only in phases with broken  gauge symmetry and signalizes about massless degrees of freedom on the string worldsheet  \cite{D'adda,Shifman}.

\section{Effective action}
\label{sec:2}
According to \cite{Itoh.Oh.Ryou1} one can build these models as a hybrid of two $\mathbb{CP}^{N-1}$ models with some coupling constants and interaction terms 
\begin{equation}\label{eq1}
L=\frac{1}{g_1^2}|D_{1\mu}\psi_1|^2+\frac{1}{g_2^2}|D_{2\mu}\psi_2|^2-i\frac{1}{\sqrt{g_1 g_2}} C_{\mu}\psi_2^{\dagger}\partial^{\mu} \psi_1-i\frac{1}{\sqrt{g_1 g_2}} C_{\mu}^*\psi_1^{\dagger}\partial^{\mu} \psi_2+\frac{1}{4}\Big(\frac{g_1}{g_2}+\frac{g_2}{g_1}\Big)C^*_{\mu}C^{\mu}
\end{equation}
where $\psi_1$ and $\psi_2$ are two complex N-vectors, such that $\psi_i^{\dagger}\psi_j=\delta_{ij}$ ($i=1,2$), coupled to the Abelian gauge fields $A_{\mu}$ and $B_{\mu}$ via covariant derivatives $D_{1\mu}=\partial_{\mu}-iA_{\mu}$ and $D_{2\mu}=\partial_{\mu}-iB_{\mu}$.
 If one starts from different coupling constants for each complex space this leads to the flag manifold.  The Grassmannian manifold corresponds to the case with equal coupling constants. Note that the presence of two different coupling constants leads to the generation of two different masses.

It is convenient to rewrite the Lagrangian in terms of the $N\times2$ matrix $Z_{i\alpha}=(\psi_{1\alpha},\psi_{2\alpha})$ ($\alpha=1,...,N$)
\begin{equation}\label{eq2}
L=Tr((D_{\mu}Z)^\dagger(D^{\mu}Z)-\lambda(Z^{\dagger}Z-R))
\end{equation}
where we have introduced matrix 
$ R = \left( \begin{array}{cc}
 r_1 & 0\\
0 & r_2  \end{array} \right)$
with $r_1=N/g_1^2$ and $r_2=N/g_2^2$ and matrix of Lagrange multipliers
$\lambda = \left( \begin{array}{ccc}
\lambda_1& \lambda_3 \\
\lambda_3^* & \lambda_2   \end{array} \right)$ imposing the constraints $(\psi_{i\alpha})^{\dagger}\psi_{j\alpha}=\delta_{ij}r_j$.
Covariant derivative is
$D_{\mu}Z=\partial_{\mu}Z-Z\tilde A_{\mu}$ with $\tilde A_{\mu}=-i\left( \begin{array}{ccc}
\ A_{\mu} & \frac{1}{2}C_{\mu} \\
\ \frac{1}{2}C_{\mu}^* & B_{\mu} \end{array} \right)$.
At classical level the gauge fields can be eliminated by their equations of
motions. Thus, for convenience they are equaled to zero in the effective action below.

Taking into account that $\lambda$ transforms in the adjoint representation of the U(2) gauge group one can see that in the case of $r_1=r_2$ the Lagrangian \eqref{eq2} is invariant under the local $U(2)$ transformation. Vice versa, when $r_1\neq r_2$ the local symmetry is $U(1)_A\times U(1)_B$.
As well known, a linear model with $G_{global}\times H_{local}$ is gauge equivalent to the non-linear sigma model corresponding to coset space $G/H$ (see \cite{Cremmer}). So it gives us $Gr(N,2)$ and $F(N,2,1)$ manifolds.
In order to perform Gaussian integration in the partition function, let us represent the Lagrangian in terms of the $2N \times 2N$ matrix 
\begin{equation}
L=(\psi_1^{\dagger},\psi_2^{\dagger})(M^T \otimes I)\left( \begin{array}{ccc}
\psi_1 \\
\psi_2   \end{array} \right)+r_1\lambda_1+r_2\lambda_2
\end{equation}
where $M=\left( \begin{array}{ccc}
\ -\partial^2 +\lambda_1 & \lambda_3 \\
\ \lambda_3^* &  -\partial^2 +\lambda_2  \end{array} \right)$  and $I_{N\times N}$ is unit matrix.
Integrating out all but two fields $\sigma_1$ and $\sigma_2$ (the first components of vectors $\psi_1$ and $\psi_2$ respectively), that will get non-zero vacuum expectation values (VEVs),  we get the effective action
\begin{equation}\label{eq4}
S_{eff}=2NTr\log\left( \begin{array}{ccc}
\ -\partial^2 +\lambda_1 & \lambda_3 \\
\ \lambda_3^* &  -\partial^2 +\lambda_2  \end{array} \right)+\int d^2x[(\partial_{\mu}\sigma_1)^2+\lambda_1(|\sigma_1|^2-r_1)\nonumber
\end{equation}
\begin{equation}
+(\partial_{\mu}\sigma_2)^2+\lambda_2(|\sigma_2|^2-r_2)+\lambda_3 \sigma_1^* \sigma_2+\lambda_3^* \sigma_1 \sigma_2^*]
\end{equation}
Here we have changed $N \rightarrow{} 2N+1$ for convenience.
Note that the main difference between this action and the one in $\mathbb{CP}^{N-1}$ model with two condensates (see \cite{Bolognesi1}) is the presence of the off-diagonal elements $\lambda_3$, $\lambda_3^*$ that mix two sectors $\psi_1$ and $\psi_2$.
As it will be seen below, this fact leads to crucial consequences for the phase structure.

In order to calculate partition function in large-N limit we should find saddle-points of the action. We will consider only time independent solutions since time-translation symmetry is unbroken.
\section{Gap equations}
\label{sec:3}
From Eq. \eqref{eq4} it follows that the total energy of the system is the sum of the energy of the quantum fluctuations and the energy of the condensates $\sigma_1$ and $\sigma_2$ (\cite{Bolognesi2})
\begin{equation}
E=2N\sum_{n}\omega_n+\int_0^{L} dx[(\partial_{\mu}\sigma_1)^2+\lambda_1(|\sigma_1|^2-r_1)+(\partial_{\mu}\sigma_2)^2+\lambda_2(|\sigma_2|^2-r_2)+\lambda_3 \sigma_1^* \sigma_2+\lambda_3^* \sigma_1 \sigma_2^*]
\end{equation}
where $\omega^2_n$ are the eigenvalues of the problem
\begin{eqnarray}\label{eq6}
\left( \begin{array}{ccc}
\ -\partial_x^2 +\lambda_1 & \lambda_3 \\
\ \lambda_3^* &  -\partial_x^2 +\lambda_2  \end{array} \right) \left( \begin{array}{ccc}
\ f_{1,n} \\
\ f_{2,n}  \end{array} \right)=\omega_n^2 \left( \begin{array}{ccc}
\ f_{1,n} \\
\ f_{2,n}  \end{array} \right)
\end{eqnarray}
 
Varying the total energy with respect to the Lagrange multipliers we get the following saddle-point equations
\begin{eqnarray}
N\sum_n\frac{1}{\omega_n}\left( \begin{array}{ccc}
\ \frac{|f_{1,n}|^2}{\theta_1} &\frac{f_{1,n}f_{2,n}^*}{\theta_2}  \\
\ \frac{f_{1,n}^*f_{2,n}}{\theta_1} &  \frac{|f_{2,n}|^2}{\theta_2}  \end{array} \right)+\left( \begin{array}{ccc}
\ |\sigma_1|^2 &\sigma_1^*\sigma_2  \\
\ \sigma_1\sigma_2^* & |\sigma_2|^2 \end{array} \right)
-\left( \begin{array}{ccc}
\ r_1 &0  \\
\ 0 & r_2 \end{array} \right)=0
\end{eqnarray}
where we have used the normalization
\begin{eqnarray}
\theta_1=\int_{0}^L dx|f_{1,n}(x)|^2,\;\;\;\;\;\theta_2=\int_{0}^L dx|f_{2,n}(x)|^2,\;\;\;\;\;\theta_1+\theta_2=2
\end{eqnarray}
The off-diagonal elements are complex conjugate to each other, thus
$\theta_1=\theta_2=1$.
Variation with respect to $\sigma_i$ gives us the equations of motion
\begin{eqnarray}
\left( \begin{array}{ccc}
\ -\partial_x^2 +\lambda_1 & \lambda_3 \\
\ \lambda_3^* &  -\partial_x^2 +\lambda_2  \end{array} \right) \left( \begin{array}{ccc}
\ \sigma_1 \\
\ \sigma_2  \end{array} \right)= \left( \begin{array}{ccc}
\ 0 \\
\ 0  \end{array} \right)
\end{eqnarray}

 We limit ourselves to considering only real homogeneous condensates, so the saddle-point equations turn out to be as follow
\begin{eqnarray}\label{eq10}
N\sum_n\frac{1}{\omega_n}\left( \begin{array}{ccc}
\ f_{1,n}^2 &f_{1,n}f_{2,n}  \\
\ f_{1,n}f_{2,n} &  f_{2,n}^2  \end{array} \right)+\left( \begin{array}{ccc}
\ \sigma_1^2 &\sigma_1\sigma_2  \\
\ \sigma_1\sigma_2 & \sigma_2^2 \end{array} \right)
-\left( \begin{array}{ccc}
\ r_1 &0  \\
\ 0 & r_2 \end{array} \right)=0
\end{eqnarray}
and
\begin{eqnarray}\label{eq11}
\left( \begin{array}{ccc}
\ \lambda_1 & \lambda_3 \\
\ \lambda_3 &  \lambda_2  \end{array} \right) \left( \begin{array}{ccc}
\ \sigma_1 \\
\ \sigma_2  \end{array} \right)= \left( \begin{array}{ccc}
\ 0 \\
\ 0  \end{array} \right)
\end{eqnarray}
As will be shown in the next section the qualitative behavior of the solutions of these equations strongly depends on  the parameters $r_1/r_2$, $\lambda_3$ and boundary conditions.
\section{Phase structure}
\label{sec:4}
First of all, note that for the models with $L\rightarrow{}\infty$ there is a unique confinement phase with the dimensional parameters $m_i\equiv \Lambda_i=\Lambda_{uv}\exp(-2\pi/Ng_i^2)$ ($i=1,2$) and zero VEVs of $\psi_i$. As was discussed in \cite{Itoh.Oh.Ryou2} for unbounded theory the solution exists only for $\lambda_3=0$.
For the models on the finite interval with $\lambda_3=0$ from \eqref{eq10} and \eqref{eq11} it follows
\begin{equation}\label{eq12}
N\sum_n\frac{f^2_{1,n}}{\omega_n}+\sigma_1^2
-r_1=0,\;\;\;\lambda_1\sigma_1=0
\end{equation}
\begin{equation}
N\sum_n\frac{f^2_{2,n}}{\omega_n}+\sigma_2^2
-r_2=0,\;\;\;\lambda_2\sigma_2=0\label{eq13}
\end{equation}
\begin{equation}
\sigma_1\sigma_2=0\label{eq14}
\end{equation}
where $r_i=\frac{2N}{\pi}\log(\Lambda_{uv}/\Lambda_i)$. For the last equation we used the condition $\sum_n\frac{f_{1,n}f_{2,n}}{\omega_n}=0$ that is satisfied for the diagonal operator in \eqref{eq6}.
Let us impose ``mixed'' DN-ND boundary conditions:
\begin{equation}\label{eq1020}
\psi_{i\alpha}(0)=0,\;\;\;\partial_x\psi_{i\alpha}(L)=0,\;\;\;\text{if}\;\;\;\alpha=2,...,N+1;
\end{equation}
\begin{equation}\label{eq1021}
\partial_x\psi_{i\alpha}(0)=0,\;\;\; \psi_{i\alpha}(L)=0,\;\;\;\text{if}\;\;\;\alpha=N+2,...,2N+1;
\end{equation}
\begin{equation}\label{eq1022}
\partial_x\sigma_i(0)=\partial_x\sigma_i(L)
\end{equation}
In this way the first terms in \eqref{eq12} and \eqref{eq13} have the form
\begin{equation}
\Bigg(\frac{N}{2}\sum_n\frac{f^2_{i,n}}{\omega_n}\Bigg)_{DN}+\Bigg(\frac{N}{2}\sum_n\frac{f^2_{i,n}}{\omega_n}\Bigg)_{ND}
\end{equation}
It was shown by Milekhin in \cite{Milekhin.2} that the equations \eqref{eq12} and \eqref{eq13} with boundaries \eqref{eq1020}-\eqref{eq1022} have the solutions for all $L$ with Higgs-Coulomb phase transitions in $L_1\approx 1/\Lambda_1$ and $L_2\approx 1/\Lambda_2$ respectively that is depicted schematically in figure~\ref{fig:i}. However, now, due to the additional condition $\sigma_1\sigma_2=0$, we have the solution only on the region $(L_1,+\infty)$. For $L\in(L_1,L_2)$ there is the massless non-zero field $\sigma_2$ that breaks $U(1)_A\times U(1)_B$ gauge symmetry to  $U(1)_A$. For $L>L_2$ we are in the symmetric (Coulomb) phase. Note that for $r_1=r_2$ (i.e. Grassmannian manifold) $L_1=L_2$ and there is only the Coulomb phase.
\begin{figure}[h]\label{fig:i}
\center{\includegraphics[width=0.65\linewidth]{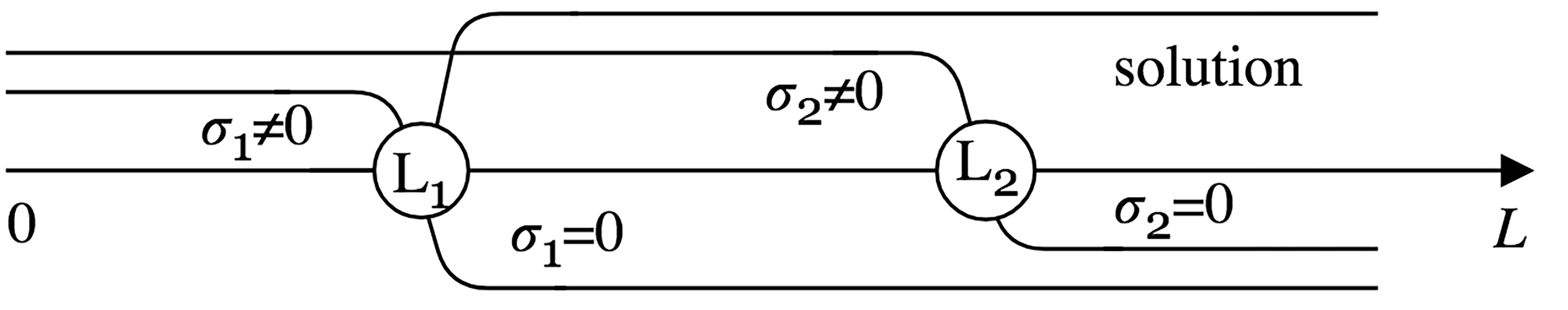}}
\caption{\label{fig:i} Schematic depiction of the phase structure. The system \eqref{eq12}-\eqref{eq14} has the solution only for $L>L_1$.}
\label{ris:image}
\end{figure}
   
Now let us examine the case of $\lambda_3=const\neq0$. The first thing we can say about this phase is that it does not exist for $L\rightarrow{}\infty$, as was  mentioned at the beginning of this section. 
To study the models on the finite interval, it is convenient to diagonalize the operator in \eqref{eq6}
\begin{equation}
\left( \begin{array}{ccc}
\ -\partial^2_x +\frac{\lambda_1+\lambda_2+\sqrt{(\lambda_1-\lambda_2)^2+4\lambda_3^2}}{2} & 0 \\
\ 0 &  -\partial^2_x +\frac{\lambda_1+\lambda_2-\sqrt{(\lambda_1-\lambda_2)^2+4\lambda_3^2}}{2}  \end{array} \right) \left( \begin{array}{ccc}
\ \tilde f_{1,n} \\
\ \tilde f_{2,n}  \end{array} \right)= \omega_n^2\left( \begin{array}{ccc}
\ \tilde f_{1,n} \\
\ \tilde f_{2,n} \end{array} \right)
\end{equation}
In this case the ``quantum'' part of the total energy has the form 
\begin{equation}
2N\Big(\sum_{n}\omega_{1,n}+\sum_{k}\omega_{2,k}\Big)
\end{equation}
where $\omega^2_{1,n}$ and $\omega^2_{2,k}$ are eigenvalues of the problems
\begin{equation}
\Big(-\partial^2_x +\frac{\lambda_1+\lambda_2+\sqrt{(\lambda_1-\lambda_2)^2+4\lambda_3^2}}{2}\Big)\tilde f_{1,n}=\omega^2_{1,n}\tilde f_{1,n}
\end{equation}
\begin{equation}
\Big(-\partial^2_x +\frac{\lambda_1+\lambda_2-\sqrt{(\lambda_1-\lambda_2)^2+4\lambda_3^2}}{2}\Big)\tilde f_{2,k}=\omega^2_{2,k}\tilde f_{2,k}
\end{equation}
After variation with respect to $\lambda_{1,2,3},\sigma_{1,2}$ and  some algebra we have the following saddle-point equations
\begin{equation}
N\sum_n\frac{\tilde f_{1,n}^2}{\omega_{1,n}}+N\sum_k\frac{\tilde f_{2,k}^2}{\omega_{2,k}}+\sigma_1^2+\sigma_2^2-r_1-r_2=0
\end{equation}
\begin{equation}\label{eq23}
\frac{(\lambda_1-\lambda_2)}{\sqrt{(\lambda_1-\lambda_2)^2+4\lambda_3^2}}\Big(N\sum_n\frac{\tilde f^2_{1,n}}{2\omega_{1,n}}-N\sum_k\frac{\tilde f^2_{2,k}}{2\omega_{2,k}}\Big)+\sigma_1^2-\sigma_2^2+r_1-r_2=0
\end{equation}
\begin{equation}\label{eq24}
\frac{2\lambda_3}{\sqrt{(\lambda_1-\lambda_2)^2+4\lambda_3^2}}\Big(N\sum_n\frac{\tilde f^2_{1,n}}{\omega_{1,n}}-N\sum_k\frac{\tilde f^2_{2,k}}{\omega_{2,k}}\Big)+2\sigma_1\sigma_2=0
\end{equation}
\begin{equation}\label{eq25}
\Big(\lambda_1-\frac{\lambda_3^2}{\lambda_2}\Big)\sigma_1=0
\end{equation}
\begin{equation}\label{eq26}
\sigma_2=-\frac{\lambda_3}{\lambda_2}\sigma_1
\end{equation}
The last two equations illustrate two possibilities: $\sigma_i$ are zero or not.
Let us first consider non-zero $\sigma_i\neq0$ that completely break initial gauge symmetry. In this case from Eq. \eqref{eq25} we get $\lambda_1\lambda_2=\lambda_3^2$. Then, using \eqref{eq23} and \eqref{eq24} we get $r_1=r_2$. 
This means that the non-zero $\lambda_3$ is compatible only with the Grassmannian model.
Finally, 
using the notations $\sigma\equiv\sigma_1=-\frac{\lambda_3}{\lambda_2}\sigma_2$ and $r\equiv r_1$ we come to the system of equations
\begin{equation}\label{eq27}
N\sum_n\frac{\tilde f_{1,n}^2}{\omega_{1,n}}-r=0;\;\;\;\;\;\;\;(-\partial^2_x +\lambda_1+\lambda_2)\tilde f_{1,n}=\omega^2_{1,n}\tilde f_{1,n}
\end{equation}
\begin{equation}\label{eq28}
N\sum_k\frac{\tilde f_{2,k}^2}{\omega_{2,k}}+\frac{\lambda_1+\lambda_2}{\lambda_2}\sigma^2-r=0;\;\;\;\;\;\;\;-\partial^2_x\tilde f_{2,k}=\omega^2_{2,k}\tilde f_{2,k}
\end{equation}
\begin{equation}\label{eq29}
\lambda_1\lambda_2=\lambda_3^2
\end{equation}

We now find out what kind of boundaries we could impose on the fields.  According to \cite{Milekhin.2}, the Eq. \eqref{eq27} has the solution on the region $L\in(\Lambda_1^{-1},+\infty)$ for DN-ND boundaries and on $(0,+\infty)$ for DD-NN boundaries on the $\tilde f_{1,n}$ modes. On the other hand, Eq. \eqref{eq28} with non-zero $\sigma$ has the solution on $(0,\Lambda_1^{-1})$ for DN-ND and has no solution for DD-NN boundaries on the $\tilde f_{2,n}$ modes ($\Lambda_1=\Lambda_2$ and is taken from the unbounded theory).                       In this way we are forced to choose the DD-NN boundaries on $\tilde f_{1,n}$ and DN-ND on $\tilde f_{2,n}$. It corresponds to the solution on the region $L\in(0,\Lambda_1^{-1})$. If $\lambda_1=\lambda_2=\lambda_3$ the  boundary conditions have the following form in terms of the fields $\psi_{i\alpha}$ 
\begin{equation}\label{eq30}
\psi_{i\alpha}(0)=0,\;\;\;\partial_x(\psi_{1\alpha}-\psi_{2\alpha})(L)=(\psi_{1\alpha}+\psi_{2\alpha})(L)=0,\;\;\;\text{if}\;\;\;\alpha=2,...,N+1;
\end{equation}
\begin{equation}\label{eq31}
\partial_x\psi_{i\alpha}(0)=0,\;\;\;\partial_x(\psi_{1\alpha}+\psi_{2\alpha})(L)=(\psi_{1\alpha}-\psi_{2\alpha})(L)=0,\;\;\;\text{if}\;\;\;\alpha=N+2,...,2N+1;
\end{equation}
\begin{equation}\label{eq32}
\partial_x\sigma_i(0)=\partial_x\sigma_i(L)
\end{equation}

Let us consider the ``classical'' part of the energy density
\begin{equation}
\lambda_1\sigma_1^2+\lambda_2\sigma_2^2+2\lambda_2\sigma_1\sigma_2-(\lambda_1+\lambda_2)r
\end{equation}
The quadratic form in this expression can be diagonalized, for example, as follows
\begin{equation}\label{eq320}
\lambda_1\sigma_1^2+\lambda_2\sigma_2^2+2\lambda_2\sigma_1\sigma_2=\lambda_2\Big(\sigma_2+\frac{\lambda_3}{\lambda_2}\sigma_1\Big)^2+\Big(\lambda_1-\frac{\lambda_3^2}{\lambda_2}\Big)\sigma_1^2
\end{equation}
Using the Eqs. \eqref{eq26} and \eqref{eq29} and also Sylvester's law of inertia we can conclude that there is one massive and one massless field.
It can be note that the detection of the massless degree of freedom allows us to expect the appearance of corresponding power corrections in the total energy (see the next section~\ref{sec:5}).

Let us now investigate the case $\sigma_1=\sigma_2=0$.
As in previous case, from Eqs. \eqref{eq23} and \eqref{eq24} we get the condition $r_1=r_2=r$,
and the system of equations is
\begin{eqnarray}
N\sum_n\frac{\tilde f_{1,n}^2}{\omega_{1,n}}-r=0,\;\;\;\;\;\;\;\Big(-\partial^2_x +\frac{\lambda_1+\lambda_2+\sqrt{(\lambda_1-\lambda_2)^2+4\lambda_3^2}}{2}\Big)\tilde f_{1,n}=\omega^2_{1,n}\tilde f_{1,n}
\end{eqnarray}
\begin{eqnarray}
N\sum_k\frac{\tilde f_{2,k}^2}{\omega_{2,k}}-r=0,\;\;\;\;\;\;\;\Big(-\partial^2_x +\frac{\lambda_1+\lambda_2-\sqrt{(\lambda_1-\lambda_2)^2+4\lambda_3^2}}{2}\Big)\tilde f_{2,k}=\omega^2_{2,k}\tilde f_{2,k}
\end{eqnarray}
This implies an equality
\begin{eqnarray}
N\sum_n\frac{\tilde f_{1,n}^2}{\omega_{1,n}}=N\sum_k\frac{\tilde f_{2,k}^2}{\omega_{2,k}}
\end{eqnarray}
that is not valid for the translation-invariant $\lambda_1$ and $\lambda_2$. Thus, 
the phase with zero $\sigma_i$ and non-zero $\lambda_3$ is absent.

We have studied $Gr(N,2)$ and $F(N,2,1)$ NLSM on the interval and find that for $\lambda_3=0$ the length of the interval $L$ is bounded from below, vice versa,  for $\lambda_3\neq0$ $L$ is bounded from above. It is now very simple to build a model that lives on a region of $L$ that bounded from below and above. Indeed, let us consider NLSM on $F(N,3,1)$ manifold with the following effective action 
\begin{equation}
S_{eff}=(N-1)Tr\ln\left( \begin{array}{ccc}
\ -\partial^2 +\lambda_1 & \lambda_{12} & \lambda_{13} \\
\ \lambda_{12} &  -\partial^2 +\lambda_2 & \lambda_{23} \\
\ \lambda_{13} & \lambda_{23} &  -\partial^2 +\lambda_3\end{array} \right)+\int d^2x[(\partial_{\mu}\sigma_1)^2+\lambda_1(|\sigma_1|^2-r_1)\nonumber
\end{equation}
\begin{equation}
+(\partial_{\mu}\sigma_2)^2+\lambda_2(|\sigma_2|^2-r_2)+(\partial_{\mu}\sigma_3)^2+\lambda_3(|\sigma_3|^2-r_3)+2\lambda_{12} \sigma_1 \sigma_2+2\lambda_{13} \sigma_1 \sigma_3+2\lambda_{23} \sigma_2 \sigma_3]
\end{equation}
and suppose that $\lambda_{12}=\lambda_{13}=0,$ $\lambda_{23}\neq0$ and $r_3=r_2>r_1$ in gap equations. One can easy to make sure that it corresponds to a phase that lives on a region of $L\in(\Lambda_1^{-1},\Lambda_2^{-1})$, where  $\Lambda_i=\Lambda_{uv}\exp(-2\pi/Ng_i^2)$ ($i=1,2,3$) and $g_2=g_3$.

\section{L-dependence}
\label{sec:5}
We now investigate L-dependance of the total energy for the different phases. Firstly consider the flag model (see figure~\ref{fig:i}). Performing calculations parallel to the work \cite{Milekhin.1} (see Appendix \ref{A}) we get the following expressions
\begin{equation}
E(L)=\frac{NLm_1^2}{\pi}\sum_{n=1}^{+\infty}(-1)^{n+1}\frac{K_1(2Lm_1n)}{Lm_1n}+\frac{NLm_1^2}{4\pi}+(m_1\leftrightarrow m_2),\;\;\;\text{if}\;\;\; L>L_2,\;\;\; m_iL>>1
\end{equation}
\begin{equation}\label{eq340}
E(L)=\frac{NLm_1^2}{\pi}\sum_{n=1}^{+\infty}(-1)^{n+1}\frac{K_1(2Lm_1n)}{Lm_1n}+\frac{NLm_1^2}{4\pi}+\frac{N\pi}{96L},\;\;\;\text{if}\;\;\;L_2>L>L_1,\;\;\; m_1L>>1
\end{equation}
where $K_1$ is modified Bessel function, $m_1=\Lambda_1$ and $m_2=\Lambda_2$. The first terms in both equations correspond to the Casimir energy for a massive complex scalar field with ``mixed'' DN boundary conditions and are negligible if $m_1L>>1$ . The L\"{u}sher term in the Higgs phase $\frac{N\pi}{96L}$ is merely the massless limit of the first term. One can associate its appearance precisely with the presence of a massless degree of freedom $\sigma_2$ in     this phase. Note that the boundary terms $m_i/2$ in the Casimir energy and  the $1/L$ corrections of masses are absent for these boundary conditions. The same is for the anomaly term \cite{Betti}.
         
For the Grassmannian model in the phase with $\lambda_3\neq0$ and completely broken $U(2)$ gauge symmetry the total energy is
\begin{eqnarray}\label{eq3400}
E(L)=\frac{NLm^2}{4\pi}-\frac{N\pi}{8L}+\frac{N\pi}{96L},\;\;\;\text{if}\;\;\; L<L_1,\;\;\; mL>>1
\end{eqnarray}
where $m^2=\Lambda_1^2+\Lambda_2^2$. As was deduced in the previous section (see Eq. \eqref{eq320}) there is one massless field, thus the term $\frac{N\pi}{96L}$ appears.

\section{Conclusion}
\label{sec:6}

We have studied the homogeneous solutions for 2d NLSMs of the Grassmannian $Gr(N,2)$ and the flag $F(N,2,1)$ manifolds on the interval and found out the strong interdependence between the phase structure and the boundary conditions. The flag model with the ``mixed'' DN boundaries may undergo the phase transition from the phase with $U(1)\times U(1)$ gauge symmetry to the phase with $U(1)$. However, for $L<\Lambda_1^{-1}$ there is no solution for these boundaries. On the other hand, for the Grassmannian model the Robin boundaries (\ref{eq30}) are compatible only with the phase with completely broken $U(2)$ gauge symmetry and $L<\Lambda_1^{-1}$. 
 
 The generalizations to other boundary conditions can be made:\\
 \RNumb{1}.   For the systems $F(N,2,1)$ and $Gr(N,2)$ on the interval the minimal length $L$ appears if any of the following boundary conditions are imposed on the fields $\psi_1$ and $\psi_2$: ``mixed'' DN-ND, periodic and ``unmixed'' DD, NN. Indeed, the $\mathbb{CP}^{N-1}$ model with listed boundaries gets non-zero VEV for a sufficiently small $L$ (see \cite{Milekhin.2}, \cite{Monin} and \cite{Bolognesi2} respectively). Then, Eq. \eqref{eq14} tells us that the saddle-point equations have no solution in this region. The generalization for the maximal length seems to be more subtle and is the subject of further study.\\
\RNumb{2}. The systems $F(N,2,1)$ and $Gr(N,2)$ have no completely deconfined phase, that can be deduced from \RNumb{1} and Eq. \eqref{eq29}\footnote{After the first version of this paper, the paper \cite{Betti.Bolognesi} appeared in arXiv. Among other things, it was argued that $\mathbb{CP}^{N-1}$ with twisted b.c. $n_k(x)=e^{i2\pi k/N}n_k(x+L), k=1,...,N,$ have a unique confinement phase with dynamical generation of the mass gap. It allows one to generalize the statement \RNumb{2} to these boundary conditions.}.

Also, it seems interesting to consider a model with general flag manifold. We expect it to have more complicated phase structure. For example, 2d sigma model with the complete flag $SU(N)/U(1)^{N-1}$ target space was recently considered in \cite{Tanizaki,Ohmori} and it was argued that the theory could be gapless in the infrared limit. Another point worth researching is the construction of inhomogeneous solutions for the Grassmannian and flag NLSMs.
\section*{Acknowledgments}
The author is grateful to A. S. Gorsky for suggestion this problem and for many valuable comments. This work was partially supported by the RFBR grant No. 19-02-00214A.

\appendix
\section{Derivation of L-dependence}\label{A}
Let us hold a derivation of Eq. \eqref{eq340} for total energy L-dependence in the phase with partially broken symmetry.    
It corresponds 
$$\lambda_3=0,\;\;\;m_2=0,\;\;\;\sigma_2\neq0,\;\;\;m_1\neq0,\;\;\;\sigma_1=0$$
therefor the gap Eq. \eqref{eq12} with ``mixed'' DN b.c. is
\begin{eqnarray}
N\sum_{n=1}^{\infty}\int_{-\infty}^{\infty}\frac{dk}{2\pi L}\frac{1}{k^2+\Big(\frac{\pi(n-1/2)}{L}\Big)^2+\lambda_1}-r_1=0
\end{eqnarray}
One can calculate the sum using the following formulas
\begin{eqnarray}
\sum_{\mathbb{Z}}^{}\frac{1}{\Big(\frac{\pi n}{L}\Big)^2+\omega^2}=\frac{2L}{\omega}\Big(\frac{1}{2}+\frac{1}{\exp(2L\omega)-1}\Big)
\end{eqnarray}
and
\begin{eqnarray}
2\sum_{\mathbb{N}}^{}\frac{1}{\Big(\frac{\pi(n-1/2)}{L}\Big)^2+\omega^2}=\sum_{\mathbb{Z}}^{}\frac{1}{\Big(\frac{\pi n}{2L}\Big)^2+\omega^2}-\sum_{\mathbb{Z}}^{}\frac{1}{\Big(\frac{\pi n}{L}\Big)^2+\omega^2}
\end{eqnarray}
Thus we have
\begin{equation}\nonumber
r_1-\frac{N}{2\pi L}\int_0^{+\infty}dk\Bigg[\frac{L}{\sqrt{k^2+m_1^2}}+\frac{4L}{\sqrt{k^2+m_1^2}}\frac{1}{\exp(4L\sqrt{k^2+m_1^2}-1)}
\end{equation}
\begin{equation}
-\frac{2L}{\sqrt{k^2+m_1^2}}\frac{1}{\exp(2L\sqrt{k^2+m_1^2}-1)}\Bigg]=0
\end{equation}
Let us introduce a quantity
\begin{eqnarray}
Q(m_1L)=\int_0^{\infty}\frac{2dk}{\sqrt{k^2+(m_1L)^2}}\frac{1}{\exp(2\sqrt{k^2+(m_1L)^2})-1}
\end{eqnarray}
and use the equation
$\Lambda_1=\Lambda_{uv}\exp(-2\pi/Ng_1^2)$
then come to the following equalities
\begin{eqnarray}
2\pi r_1-N\log(\Lambda_{uv}L)=-N\log(\Lambda_1 L)=-N\log(m_1L)+2NQ(2m_1L)-NQ(m_1L)
\end{eqnarray}
If  $m_1L>>1$
\begin{eqnarray}\label{eq3200}
Q(m_1L)=\frac{\sqrt{\pi/m_1}e^{-2m_1L}}{\sqrt{L}}
\end{eqnarray}
Therefor $m_1=\Lambda_1$
and 1/L corrections are absent.
\\




After Pauli-Villars regularization effective action is
\begin{equation}
S_{eff}^{reg}=2N\sum^{2}_{i=0}c_iTr\log(-\partial^2+m_1^2+\tilde m_i^2)-\int d^2xr_1m_1^2+...
\end{equation}
where ``...'' means $m_1$-independent terms, and
\begin{equation}
\tilde m_0=0,\;\;\;c_0=1,\;\;\:c_1=\frac{\tilde m_2^2}{\tilde m_1^2-\tilde m_2^2},\;\:\:c_2=\frac{-\tilde m_1^2}{\tilde m_1^2-\tilde m_2^2}
\end{equation}
Varying it with respect to $m_1^2$, we get
\begin{equation}
\int dx_0r_1=N\sum^{2}_{i=0}c_iTr\frac{1}{-\partial^2+m_1^2+\tilde m_i^2}
\end{equation}
If $m_1L>>1$ then the approximation \eqref{eq3200} is valid, and
\begin{equation}
r_1=\frac{1}{2\pi L}\Bigg(\frac{L}{2}\log\bigg(\frac{m_1^2+\tilde m_2^2}{m_1^2}\bigg)+\frac{L\tilde m_2^2}{2(\tilde m_1^2-\tilde m_2^2)}\log\bigg(\frac{m_1^2+\tilde m_2^2}{m_1^2+\tilde m_1^2}\bigg)\Bigg)
\end{equation}
Suppose that $\tilde m_1^2=aM^2$ and $\tilde m_2^2=M^2$ and take 
\begin{eqnarray}
a\rightarrow{1},\;\;\;M\rightarrow{\infty}
\end{eqnarray}
then regularised value for $r_1$ is
\begin{eqnarray}
r_1^{reg}=-\frac{N}{4\pi}
\end{eqnarray}

The quantity $Tr\log(-\partial^2+m_1^2)$ is the Casimir energy for a massive scalar field. For mixed DN-ND boundary conditions, it is
\begin{eqnarray}
\frac{Lm_1^2}{\pi}\sum_{n=1}^{+\infty}(-1)^{n+1}\frac{K_1(2Lm_1n)}{Lm_1n}
\end{eqnarray}
The quantity $Tr\log(-\partial^2+m_2^2)$, where $m_2=0$, is the Casimir energy for a massless scalar field
\begin{eqnarray}     
\frac{N\pi}{96L}
\end{eqnarray}
Putting it all together we get
\begin{equation}
E(L)=\frac{NLm_1^2}{\pi}\sum_{n=1}^{+\infty}(-1)^{n+1}\frac{K_1(2Lm_1n)}{Lm_1n}+\frac{NLm_1^2}{4\pi}+\frac{N\pi}{96L},\;\;\;\text{if}\;\;\;L_2>L>L_1,\;\;\; m_1L>>1
\end{equation}
Similarly one can get Eq. \eqref{eq3400}. 


\end{document}